\documentclass[preprint,3p,times,twocolumn]{elsarticle}

\usepackage{graphics}
\usepackage{graphicx}
\usepackage{amssymb}
\usepackage{subfigure}
\usepackage{marvosym}
\usepackage{amsmath}
\usepackage{multirow}
\usepackage{fancyhdr}
\usepackage{rotate}

\usepackage{lineno} 

\renewcommand{\d}{\mathrm{d}}
\newcommand{\pt}{p_\mathrm{T}}

\journal{Physics Letter B}

\begin{document}

\begin{frontmatter}

\title{Precision study of the $\eta\to\mu^+\mu^-\gamma$ and $\omega\to\mu^+\mu^-\pi^0$ electromagnetic
      transition form-factors and of the $\rho\to\mu^+\mu^-$ line shape in NA60} 

\author[1]{R.~Arnaldi}
\author[2,3]{K.~Banicz}
\author[4]{K.~Borer}
\author[5]{J.~Castor}
\author[6]{B.~Chaurand}
\author[7]{W.~Chen}
\author[8]{C.~Cical\`o}
\author[1]{A.~Colla}
\author[1]{P.~Cortese}
\author[2,3]{S.~Damjanovic}
\author[2,9]{A.~David} 
\author[8]{A.~de~Falco}
\author[5]{A.~Devaux}
\author[10]{L.~Ducroux} 
\author[11]{H.~En'yo}
\author[5]{J.~Fargeix}
\author[1]{A.~Ferretti}
\author[8]{M.~Floris}
\author[2]{A.~F\"orster}
\author[5]{P.~Force}
\author[2,5]{N.~Guettet}
\author[10]{A.~Guichard}
\author[12]{H.~Gulkanian} 
\author[11]{J.~M.~Heuser}
\author[2]{P.~Jarron}
\author[2,9]{M.~Keil} 
\author[6]{L.~Kluberg}
\author[7]{Z.~Li}
\author[2]{C.~Louren\c{c}o}
\author[9]{J.~Lozano} 
\author[5]{F.~Manso} 
\author[2,9]{P.~Martins}
\author[8]{A.~Masoni}
\author[9]{A.~Neves}
\author[11]{H.~Ohnishi}
\author[1]{C.~Oppedisano}
\author[2,9]{P.~Parracho}
\author[10]{P.~Pillot}
\author[12]{T.~Poghosyan}
\author[8]{G.~Puddu}
\author[2]{E.~Radermacher}
\author[2,9]{P.~Ramalhete} 
\author[2]{P.~Rosinsky}
\author[1]{E.~Scomparin}
\author[9]{J.~Seixas}
\author[8]{S.~Serci} 
\author[2,9]{R.~Shahoyan} 
\author[9]{P.~Sonderegger}
\author[3]{H.~J.~Specht}
\author[10]{R.~Tieulent}
\author[8,10]{A.~Uras\corref{cor1}}
\author[8]{G.~Usai\corref{cor1}}
\author[9]{R.~Veenhof}
\author[2,8,9]{H.~K.~W\"ohri}
\author[]{\\(NA60 Collaboration)}

\address[1]{Universit\`a di Torino and INFN, Italy}
\address[2]{CERN, Geneva, Switzerland}
\address[3]{Physikalisches Institut der Universit\"{a}t Heidelberg, Germany}
\address[4]{University of Bern, Switzerland}
\address[5]{LPC, Universit\'e Blaise Pascal and CNRS-IN2P3, Clermont-Ferrand, France}
\address[6]{LLR, Ecole Polytechnique and CNRS-IN2P3, Palaiseau, France}
\address[7]{BNL, Upton, NY, USA}
\address[8]{Universit\`a di Cagliari and INFN, Italy}
\address[9]{IST-CFTP, Lisbon, Portugal}
\address[10]{IPN-Lyon, Univ.\ Claude Bernard Lyon-I and CNRS-IN2P3, Lyon, France}
\address[11]{RIKEN, Wako, Saitama, Japan}
\address[12]{YerPhI, Yerevan, Armenia}      

\cortext[cor1]{~Corresponding authors: antonio.uras@cern.ch (A.~Uras), \\ gianluca.usai@ca.infn.it (G.~Usai)} 

\begin{keyword}
Lepton Pairs\sep{Transition form factor}\sep{Conversion decays}\sep{Rho meson}
 \PACS{13.85.Qk}\sep{13.40.Gp}\sep{13.20.-v} 
\end{keyword}

\begin{abstract} 
The NA60 experiment studied low-mass muon pair production in
proton-nucleus (p-A) collisions using a 400~GeV proton beam at the CERN SPS. 
The low-mass dimuon spectrum is well described by the superposition of the two-body and
Dalitz decays of the light neutral mesons $\eta$, $\rho$, $\omega$,
$\eta'$ and $\phi$, and no evidence of in-medium effects is found.
A new high-precision measurement of the electromagnetic
transition form factors of the $\eta$ and $\omega$ was performed,
profiting from a 10~times larger data sample than the peripheral In-In sample previously collected by NA60.
Using the pole-parameterisation $|F(M)|^2 = (1 -M^2/\mathrm{\Lambda}^2)^{-2}$ we find
$\mathrm{\Lambda}_\eta^{-2} = 1.934\ \pm\ 0.067$~(stat.) $\pm\ 0.050$~(syst.)~(GeV/$c^2$)$^{-2}$ and 
$\mathrm{\Lambda}_\omega^{-2} = 2.223\ \pm\ 0.026$~(stat.) $\pm\ 0.037$~(syst.)~(GeV/$c^2$)$^{-2}$.
An improved value of the branching ratio of the Dalitz decay
$\omega \to \mu^+\mu^-\pi^0$ is also obtained, with $BR(\omega \to \mu^+\mu^-\pi^0) = [1.41~\pm~0.09~\mathrm{(stat.)}$
$\pm~0.15~\mathrm{(syst.)}] \times 10^{-4}$.
Further results refer to the $\rho$ line shape and a new limit on $\rho/\omega$ interference
in hadron interactions.
\end{abstract}

\end{frontmatter}


\section{Introduction}

\noindent Dimuon production in proton-nucleus (p-A) interactions at
SPS energies of 400 GeV ($\sqrt{s_\mathrm{NN}} = 27.5$~GeV), for masses below 1~GeV/$c^2$, is dominated by the two-body and Dalitz
decays of the vector mesons $\rho$, $\omega$, and $\phi$ and the
pseudoscalar meson $\eta$. Beyond serving as a reference for the observations in ultra-
relativistic heavy-ion collisions, p-A data also permit to measure important properties 
of the produced particles, essentially undisturbed by the nuclear medium due to the 
large rapidity gap between central production and the target rapidity region. The present 
paper reports on a new measurement of the electromagnetic transition form-factors of 
the $\eta$ and $\omega$ Dalitz decays, improving with a still higher precision the previous NA60 
results based on peripheral In-In collisions~\cite{Arnaldi:2009wb}. Supplementary new information is 
also obtained for the branching ratio of the $\omega$ Dalitz decay
$\omega \to \mu^+ \mu^- \pi^0$, the line shape of the $\rho$~meson 
and possible $\rho/\omega$ interference effects. 

Transition form factors are an important ingredient in the detailed understanding of the
nature of mesons and their underlying quark and gluon structure.
In this context, recent improved measurements made
it possible to set new stricter constraints for theoretical models~(see~\cite{Czerwinski:2012ry}
and references therein). 
Independent support for further improvements of form factor data arises from the impact 
on the hadronic light-by-light contributions to the anomalous magnetic
moment of the muon~\cite{Pauk:2014rta,Colangelo:2014pva}.
In addition, the previous lack of precise measurements of the form factors was one of the
main sources of systematic uncertainties in the description of low mass dilepton spectra
in heavy-ion collisions, with major implications for the study of the 
in-medium modifications of the $\rho$ meson~\cite{Agakichiev:2005ai, Specht:2007ez}.

In the Dalitz decays $\eta \to \mu^+ \mu^- \gamma$ and $\omega \to \mu^+ \mu^- \pi^0$, 
the mesons decay electromagnetically into a virtual photon with mass $M$ $-$ in turn converting 
into a lepton pair $-$ and a third body. The form factor $|F(M)|^2$ quantifying 
the deviation from the point-like behaviour in pure QED~\cite{Kroll:1955zu}, 
due to the internal electromagnetic structure of the decaying meson, 
is directly accessible by
comparing the measured invariant mass spectrum of the lepton pairs
from the Dalitz decays with the point-like QED prediction. 
The predictions of the Vector Meson Dominance (VMD) model for the form factors of the $\eta$
and $\omega$ mesons have been
tested by the Lepton-G experiment \cite{Landsberg:1986fd,Dzhelyadin:1980tj,Dzhelyadin:1980kh} with pion beams,
and recently by the NA60 experiment in In-In peripheral collisions
\cite{Arnaldi:2009wb}. The much improved results of the latter measurement
confirmed the fact that the VMD model
strongly underestimates the observed $\omega$ form factor.
It should be mentioned that the transition form factor of the $\omega$ meson has also
been measured in the complementary reaction 
$e^+ e^- \to \omega \pi^0$ in the mass region $M > m_\omega + m_{\pi^0}$~\cite{Landsberg:1986fd,Druzhinin:1984zr}.
The difficulties in describing the form factor data in the two mass regions 
on consistent theoretical grounds has already been noted in~\cite{Landsberg:1986fd}.

\noindent The precise shape of the vector meson $\rho$, the main object
of strong in-medium modifications in nuclear collisions, has more recently
been of renewed interest in hadron collisions as a benchmark for any
deviations from the vacuum shape.
The need for a Boltzmann term $\exp(-M/T)$~\cite{Knoll:1998iu} beyond the standard
description, with $T$ being an effective temperature parameter,
was experimentally confirmed for the first time by the peripheral In-In data of NA60~\cite{Arnaldi:2009wb}.
The large data sample collected by NA60 in p-A collisions 
allows now an independent measurement of the $\rho$ line shape, in
another hadronic collision system which is expected to be free from
in-medium effects (as discussed in Section~\ref{sec:rho}).\\

\noindent In addition, based on the same argument, new investigations of possible quantum interference effects between
the $\rho$ and $\omega$ mesons can be performed, 
in the presence of a hadronic initial state. The observation of such an effect has already been reported
in $e^+ e^-$ collisions in the $\pi^+\pi^-$ channel
\cite{Barkov:1985ac, Achasov:2005rg, Aubert:2009ad, Ambrosino:2008aa, Ambrosino:2010bv}, while
measurements in p-A collisions~\cite{Veenhof:1993xt,Akesson:1994mb,Naruki:2005kd} are at present not conclusive.\\

\noindent In this paper we report on high-precision results on low mass dimuon production in p-A data, 
collected by the NA60 experiment at the CERN SPS at 400~GeV. The NA60 experiment 
accumulated a large p-A data sample for six nuclear targets: Be, Cu, In, W, Pb and~U. 
Integrating over the targets, about 180\,000 $\mu^+\mu^-$ events were collected. 
This sample is almost a factor of~10 larger than the indium-indium peripheral data exploited in~\cite{Arnaldi:2009wb}. 
With this data, a comprehensive and detailed study of the production of the light neutral mesons has been performed 
in p-A collisions, providing the most precise measurement currently available for the 
electromagnetic transition form factors in the $\eta \to \mu^+\mu^-\gamma$ and $\omega \to\mu^+\mu^-\pi^0$ 
decays, together with a new measurement of the branching ratio of the Dalitz decay $\omega \to \mu^+\mu^-\pi^0$,
the study of the line shape of the $\rho$ meson and the investigation of $\rho/\omega$ quantum interference effects.

\section{Apparatus and event selection}

\noindent During the 2004 run, the NA60 experiment collected data with
a system of nine sub-targets of different nuclear spe\-ci\-es $-$ Be,
Cu, In, W, Pb and~U $-$ simultaneously exposed to an incident
400~GeV proton beam. The individual target thicknesses were chosen so
as to collect event samples of similar sizes for each nuclear
species. The total target length was 7.5\,\% of an interaction length.
The beam was delivered by the SPS with an intensity of $2\times10^9$~protons per second in 4.8~s long bursts, 
every 16.8~s.
\subsection{Apparatus description}
\noindent A general description of the NA60 apparatus can be found for
example in~\cite{Arnaldi:2008er}. Here, only the specific details
relevant to the setup used during the proton run are given.  The
produced dimuons are identified and measured by the muon spectrometer,
composed of a set of tracking stations, trigger scintillator
hodoscopes, a toroidal magnet and a hadron absorber. The angular
acceptance is $35<\theta<120$~mrad, corresponding to the pseudo-rapidity range $2.8 < \eta_\mathrm{lab} < 4.0$. 
A silicon vertex spectrometer tracks all charged particles, including the muons, before
entering the absorber. During the proton run, the main components of this spectrometer were
10 pixel planes based on
the ALICE sensors~\cite{Wyllie:1999uv} and two pixel planes based on ATLAS sensors~\cite{Alam:1999yh}.  
The two ATLAS planes can be operated with a 50~ns gate, 4 times smaller than
the one required by the ALICE planes. This is particularly important
because the hits in the ATLAS sensors can be effectively used to
reduce the interaction pile-up: only the tracks with hits in these two
planes are kept, thus discarding the ones associated to out of time
hits in the ALICE sensors. In this way, while still
remaining non negligible, with $\sim$2 beam-target interactions per
event on top of the one giving the trigger, the interaction pile-up
can be coped with by the good granularity and redundancy of the vertex
tracker.
\subsection{Event selection, background and Monte Carlo simulations}
\noindent The muon tracks reconstructed in the muon spectrometer are
extrapolated back to the target region and matched to the
tracks reconstructed in the vertex spectrometer. This is done
comparing both their angles and momenta, requiring a {\it matching}
$\chi^2$ less than 3. Once identified, the muons are refitted using
the joint information of the muon and vertex spectrometers.  These
tracks will be referred to as \emph{matched muons}.  Muon pairs of
opposite charge are then selected.  The matching technique improves
significantly the signal-to-background ratio and the dimuon mass
resolution. The latter is \mbox{30-35}~MeV/$c^2$ (depending on the target)
at the $\omega$ mass, somewhat worse than it was in the indium run \mbox{($\sim 23$~MeV/$c^2$)}~\cite{Arnaldi:2009wb}
because of the heavier absorber setup.
The small residual combinatorial background
(originating from $\pi$ and $K$ decays) is subtracted from the real
data. Its shape is estimated with an event mixing technique, while its
normalisation is established fixing the like-sign (LS) component
from the mixing to the LS component of the direct (same-event) muon
pair sample (containing
no signal from correlated pairs at the SPS energies). The background
accounts for less than $10\%$ of the integrated mass spectrum below
1.4~GeV$/c^2$. The comparison between the mixed and real samples, in turn, gives an
average uncertainty of $10\%$ at most, for both the $(++)$ and the
$(--)$ components; because of the low absolute level of the
background and because of its smooth mass profile, this uncertainty hardly
affects the results.

The background from fake track matches, which arises at high multiplicities
from the association of a muon track  to more than
one track in the vertex spectrometer with an acceptable matching
$\chi^2$, is significantly lower than the combinatorial background. 
Its contribution is negligible in the proton-nucleus data $-$ being in any case
taken into account by the overlay Monte Carlo technique adopted for the simulations, see below.
The top panel of \figurename~\ref{fig:massSpectrum} shows the final $\mu^+\mu^-$ mass
spectrum together with the combinatorial background evaluated as described above.

The electromagnetic decays of the light, neutral pseudoscalar and
vector mesons ($\eta$, $\eta'$, $\rho$, $\omega$ and $\phi$) are the
dominating processes at the lower end of the dimuon mass spectrum (below $\sim 1.2$~GeV/$c^2$),
adding to the continuous spectrum via their Dalitz decays and/or
giving rise to distinct peaks via their 2-body decays. This hadronic
decay cocktail was simulated with the NA60 Monte Carlo generator
Genesis~\cite{genesis}. The input parameters for the kinematic
distributions of the generated processes have been tuned by comparison
with the real data, by means of an iterative procedure ensuring
self-consistency to the analysis.

The rapidity distributions in the center of mass frame were
generated according to the expression $\d N/\d y \propto 1/\cosh^2
(ay)$, similar to a Gaussian of width $\sigma =
0.75/a$, where $a$ describes the empirical functional mass dependence of the width with 
values of about 0.5 and 0.75 at the masses of 0.14~GeV/$c^2$ ($\pi^0$)
and 1~GeV/$c^2$, respectively~\cite{genesis}.  This simple
parameterisation has been used by several experiments, since it
describes reasonably well existing measurements.  

The transverse momentum spectra used in the simulations are extracted from the same p-A data set at 
400~GeV on which the present paper is based. A
preliminary analysis for these measurements has appeared elsewhere~\cite{Uras:2011qs},
showing in fact strong differences to the
158~GeV regime. 
The muon angular distributions also entering the simulations are assumed
to be isotropic for the 2-body decays, while 
the angular anisotropies of the Dalitz decays, expected to be the same for
the pseudo-scalar ($\eta,~\eta'$) and vector ($\omega$) mesons~\cite{Bratkovskaya:1996nf}, 
are described by the equation~\cite{Anastasi:2016hdx}  
\begin{equation}
	f(\theta) = 1 + \cos^2\theta + \left( \frac{2m_\mu}{M} \right)^2 \sin^2\theta~,
\end{equation}
where $M$ is the mass of the virtual photon, $m_\mu$ the mass of the muon,
and $\theta$ the angle between the positive muon and the momentum of the parent meson
in the rest-frame of the virtual photon. As was explicitly verified, the form 
factor data resulting from the present analysis actually are, within their 
statistical errors, completely immune towards the character of the angular
distribution of the Dalitz decays. 
This is due to the fact that the anisotropy of the Dalitz decays 
is strongly smeared out in the laboratory frame
and practically does not affect the dimuon acceptance.

For the mass line shapes of the narrow resonances $\eta$,
$\omega$ and $\phi$, we used the modified relativistic Breit-Wigner
parameterisation, first proposed by G.J.~Gounaris and
J.J.~Sakurai~\cite{Gounaris:1968mw}, with widths and masses
taken from the PDG~\cite{Nakamura:2010zzi}. For the broad $\rho$ meson
we used the parameterisation~\cite{Knoll:1998iu} 

\begin{footnotesize}
\begin{equation}
   \frac{\d N}{\d M} \propto \frac{\sqrt{1-\frac{4m^2_\mu}{M^2}}\left( 1+\frac{2m^2_\mu}{M^2}\right)\left( 1-\frac{4m^2_\pi}{M^2}\right) ^{3/2}}
{\left( m^2_{\rho}-M^2\right)^2+m^2_\rho \mathrm{\Gamma}^2_\rho(M)} \left(MT\right)^{3/2} e^{-\frac{M}{T_\rho}}
\end{equation}
\end{footnotesize}

\noindent with a mass dependent width 

\begin{footnotesize}
\begin{equation}
\mathrm{\Gamma}_\rho(M)=\mathrm{\Gamma}_{0\rho} \frac{m_\rho}{M}\left(
\frac{M^2/4-m^2_\mu} {m^2_\rho/4-m^2_\mu}\right)^{3/2}=\mathrm{\Gamma}_{0\rho}
\frac{m_\rho}{M} \left( \frac{q} {q_0}\right)^{3}.
\end{equation}
\end{footnotesize}

\noindent The muon mass $m_\mu$ and the pion mass $m_\pi$ were 
fixed to the PDG values~\cite{Nakamura:2010zzi},
while the value of the pole mass $m_\rho$ and the temperature $T_\rho$ were optimised using the data 
themselves as discussed in Section~\ref{sec:rho}. The width
$\mathrm{\Gamma}_{0\rho}$ was set to the PDG
value~\cite{Nakamura:2010zzi}; nevertheless, its
variation has being considered as a part of the systematic tests for the
measurement of the $\omega$ form factor, see Section~\ref{sec:formFactors}.

The dimuon mass distributions of the $\eta$ and $\omega$ Dalitz decays
are described by

\begin{footnotesize}
\begin{eqnarray}
 \hspace{-0.7cm}\frac{\mathrm{d\Gamma}(\eta\to\mu\mu\gamma)}{\mathrm{d}M}  & = & 
2M \frac{2}{3} \frac{\alpha}{\pi} \frac{\mathrm{\Gamma}(\eta \rightarrow
\gamma \gamma)}{M^2} \left( 1 - \frac{M^2}{m_\eta^2} \right)^3 \left( 1 +
\frac{2m_\mu^2}{M^2} \right)\nonumber\\
&\times& \left( 1 - \frac{4m_\mu^2}{M^2} \right)^{1/2}
 \left| F_\eta(M^2) \right|^2,
\end{eqnarray}
\end{footnotesize}
\begin{footnotesize}
\begin{eqnarray}
 \hspace{-0.7cm}\frac{\mathrm{d\Gamma}(\omega\to\mu\mu\pi^0)}{\mathrm{d}M} & = &
 2M \frac{\alpha}{3\pi} 
 \frac{\mathrm{\Gamma}(\omega \rightarrow \gamma \pi^0)}{M^2} \left( 1 +
\frac{2m_\mu^2}{M^2} \right) \left( 1 - \frac{4m_\mu^2}{M^2} \right)^{1/2} \nonumber \\
 & \times &
 \left[ \left( 1 + \frac{M^2}{m_\omega^2 -m_{\pi^0}^2} \right)^2 -
\frac{4m_\omega^2 M^2}{(m_\omega^2 -m_{\pi^0}^2)^2} \right]^{3/2} 
 \left| F_\omega(M^2) \right|^2,\nonumber\\
\end{eqnarray}
\end{footnotesize}
\hspace{-0.12cm}where the $\pi_0$,~$\eta$ and $\omega$ masses are taken from the PDG tables~\cite{Nakamura:2010zzi}.
The form factors are expressed in the pole-parameterisation:
\begin{eqnarray}
\left|F_\eta(M^2)\right|^2 &=& \left(1-M^2/\mathrm{\Lambda}_\eta^{2}\right)^{-2},\\
\left|F_\omega(M^2)\right|^2&=&
\left(1-M^2/\mathrm{\Lambda}_\omega^{2}\right)^{-2},
\end{eqnarray}
implying a monotonic rise with divergence at a position not related to 
a pole of any known particle.  

\noindent The semimuonic simultaneous decays from  $D\bar D$ mesons produce
a smooth continuum with a maximum at around 1~GeV/$c^2$. They
were simulated with PYTHIA~6.4~\cite{Sjostrand:2006za}.\\

\noindent The Monte Carlo simulations were performed using the overlay
technique, which consists of superimposing a Monte Carlo generated muon
pair onto real events, in order to realistically simulate the underlying hadronic
event together with the detector specific behaviour. A real event is
read, chosen among the reconstructed data collected by the experiment,
containing a high-mass matched dimuon (within the $J/\psi$ mass
window) whose vertex is imposed to be the origin of the generated muon pair. 
Alternatively, dimuons whose vertex has the $z$-coordinate determined with an uncertainty
smaller than 3~mm were also used. This second choice, applying weaker
conditions on the vertex candidates, has been considered for
systematic checks in the analysis.  The muon pair produced in the
simulation is tracked through the NA60 apparatus, using
GEANT3~\cite{Brun:1987ma}. Starting from the ensemble of simulated and real hits, the events
in which a muon pair gave rise to a trigger were reconstructed using
the same reconstruction settings used for the real data.  To make the
MC simulation as realistic as possible, the MC tracks leave a signal
in a given pixel plane with a probability proportional to the plane
efficiency as estimated from the analysis of the real data.\\

\begin{figure}[t!] 
   \begin{center}
    \includegraphics[width=0.44\textwidth]{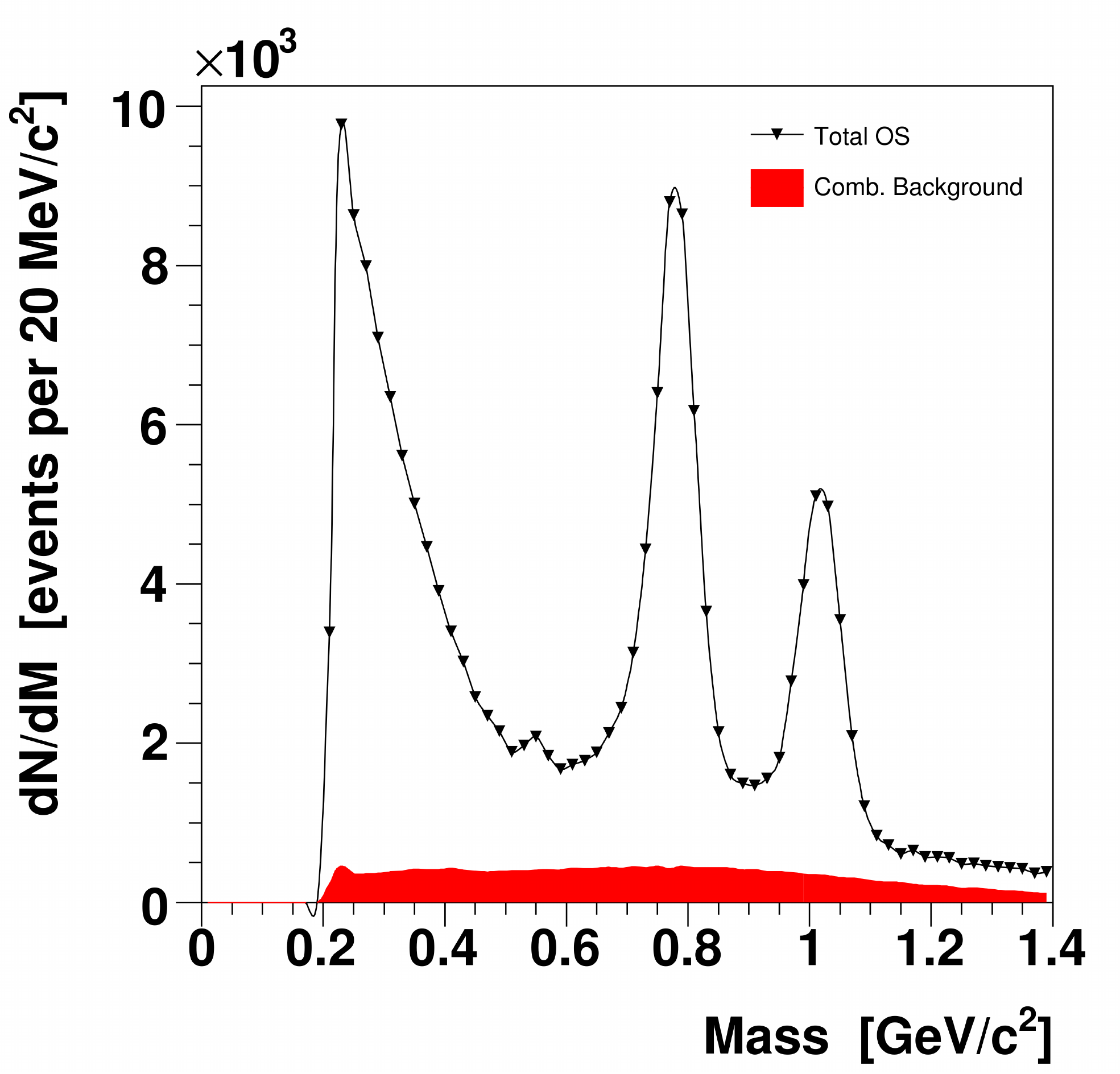}
    \includegraphics[width=0.44\textwidth]{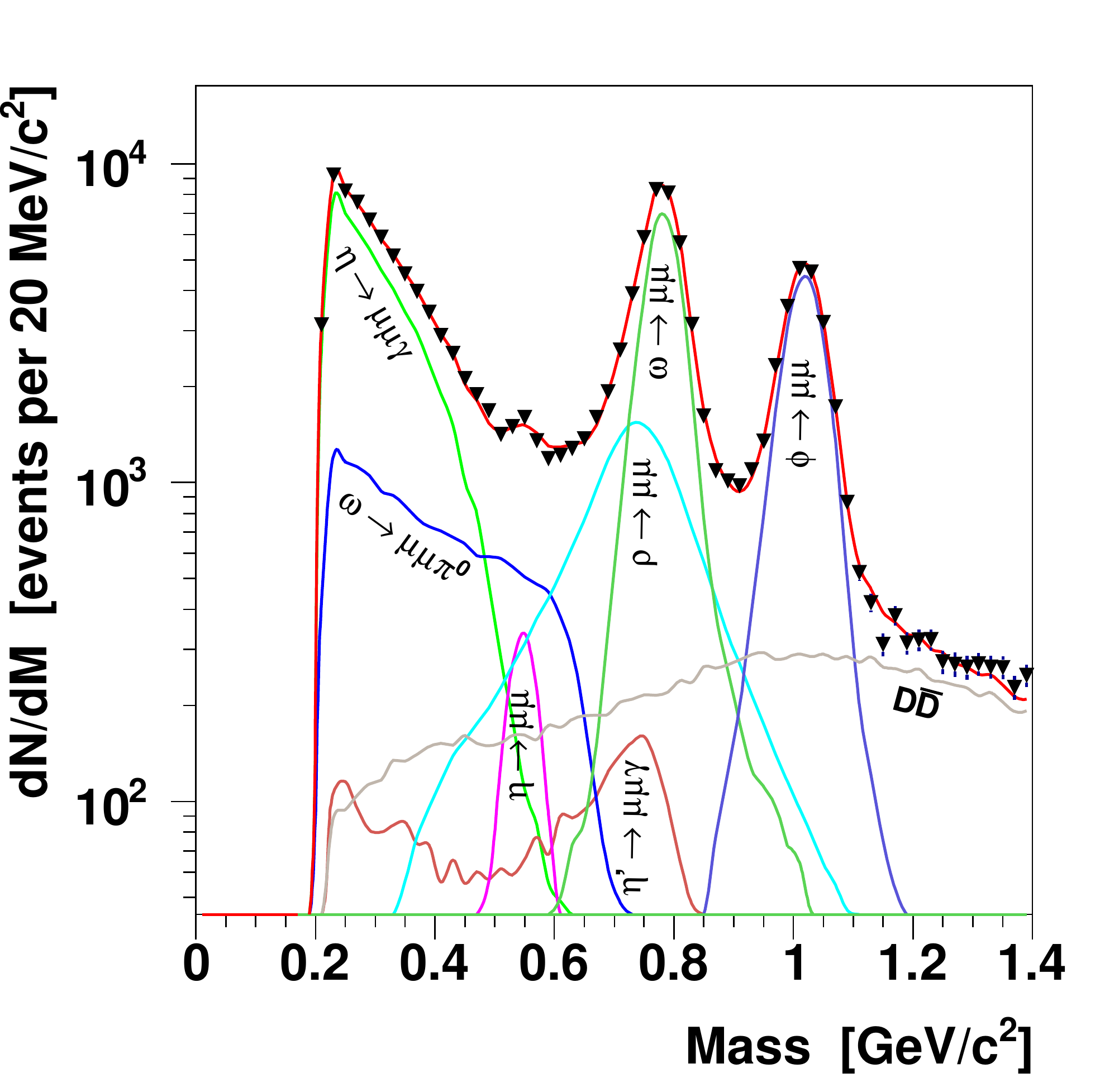} 
    \end{center} 
    \vspace{-0.2cm}
    \caption[\textwidth]{Top panel: target-integrated raw mass
spectrum and combinatorial background. Bottom panel: target-integrated mass spectrum
after subtraction of combinatorial background in comparison to the MC hadron cocktail.}
\label{fig:massSpectrum}
\vspace{0.5cm}
\end{figure}

\section{Analysis and results}

\noindent All the results presented in this paper are obtained through an iterative analysis
of the dimuon mass spectrum, defined as the signal resulting after subtraction of 
the combinatorial background. The bottom panel of \figurename~\ref{fig:massSpectrum} shows 
how the $\mu^+\mu^-$ mass spectrum is described by the hadronic cocktail fit at the
last stage of the iterative procedure, with all the parameters extracted from the data 
set to their final values.

The fit performed in terms of the superposition of the MC
processes satisfactorily describes the profile of the observed mass
spectrum. Any possible $\rho/\omega$ interference effect is neglected here, 
as will be justified in Section~\ref{sec:interference}. 
The contribution of the Dalitz decay $\eta'\to
\mu^+\mu^-\gamma$ accounts for a very small fraction of the total
dimuon yield; for this reason, and because of its continuum shape
having no dominant structure apart from the broad peak at the $\rho$
mass (due to the contribution of the $\rho$ to the $\eta'$ form
factor), the fit to the reconstructed mass spectrum is not sensitive to
this contribution, and the ratio $\sigma_{\eta'} /
\sigma_\omega$ was fixed to 0.12~\cite{genesis,Becattini:2003wp}. All the other processes 
have their normalisations free. 

\subsection{$\eta$ and $\omega$ Dalitz decay transition form factors}
\label{sec:formFactors}

\noindent The parameters optimised by the fit to the dimuon mass spectrum fix the level
of each process contributing to the MC cocktail. Using these normalisations,
we now isolate the Dalitz decays of the $\eta$ and $\omega$ mesons and the two-body
decay of the $\rho$ by subtracting all the other sources. The $\rho$
is retained, even if not directly involved in the measure of
the electromagnetic form factors, in order to better
control the systematics related to the small contribution of its
low-mass tail in the mass region of interest here 
($M < 0.65$~GeV/$c^2$). The present analysis thus isolates the Dalitz decays of the $\eta$ and $\omega$ mesons
by means of an inclusive measurement of the dimuon invariant mass, without the identification of the third body. 
Nonetheless, this approach provides reliable and remarkably precise results on the transition form factors, thanks to
the large available statistics, which ensures a good control of the competing dimuon sources.

The resulting mass spectrum is
corrected for the effects of geometrical acceptance and reconstruction
efficiency~\cite{Damjanovic:2006bd, Damjanovic:2007qm}. 
In order to do so, we build a correction profile as a
function of mass, weighting the profiles obtained from the MC simulation for
each of the three processes separately, according to the observed yields in
each mass bin. 

\begin{figure}[b!] 
   \begin{center}
    \includegraphics[width=0.44\textwidth]{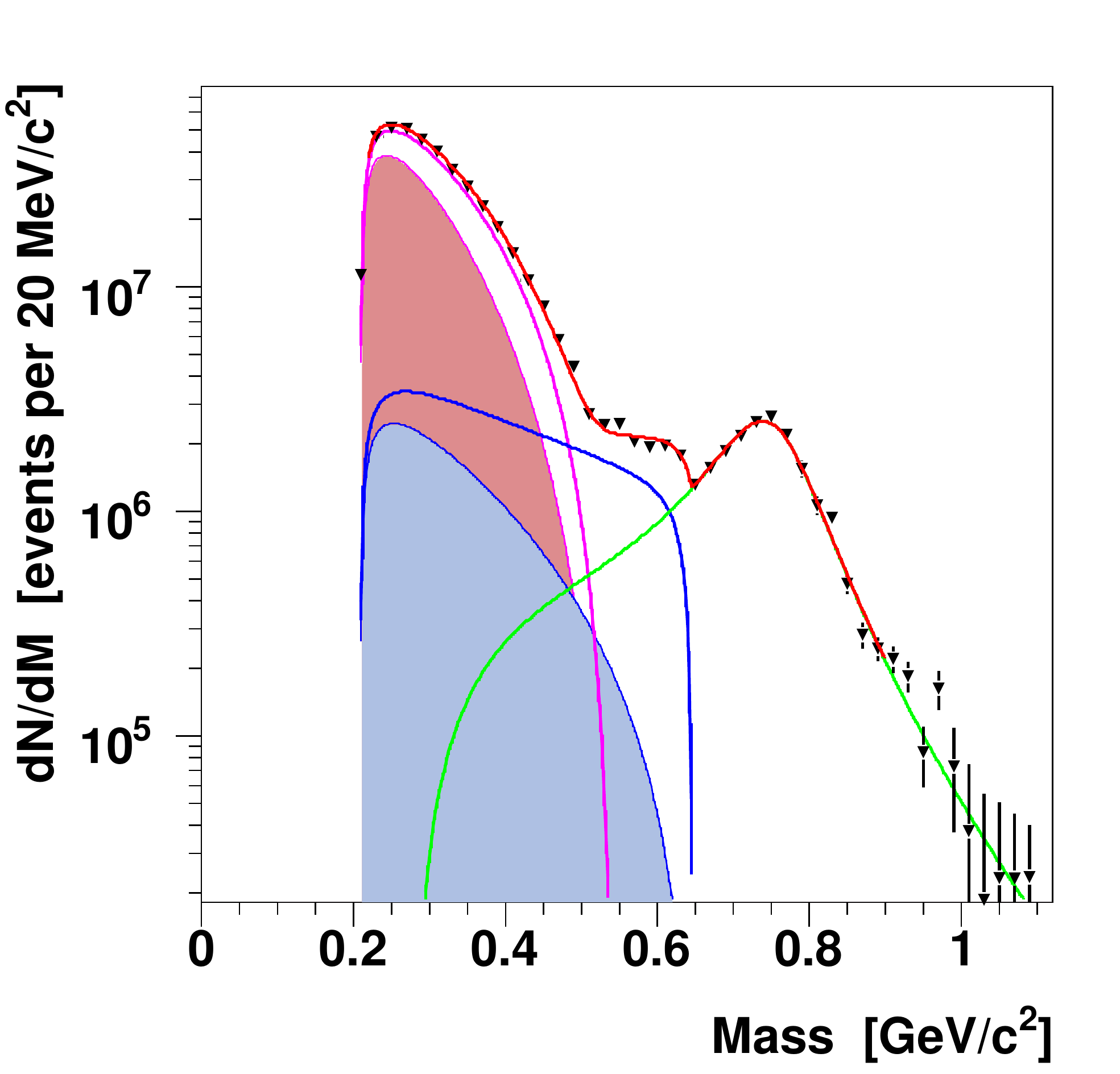} \hspace{.0\textwidth} 
    \end{center} 
\vspace{-0.1cm}
\caption[\textwidth]{Fit to the acceptance- and efficiency-corrected mass spectrum relative to the processes
$\eta\to\mu^+\mu^-\gamma$, $\omega\to\mu^+\mu^-\pi^0$ and
$\rho\to\mu^+\mu^-$. The shaded areas indicate the Kroll-Wada
expectations for point-like particles, defined by
QED~\cite{Kroll:1955zu}.}
\label{fig:FormFactorsFit}
\end{figure}

\begin{figure}[htbp] 
   \begin{center}
    \vspace{-0.5cm}
    \hspace{-0.04\textwidth} \includegraphics[width=0.44\textwidth]{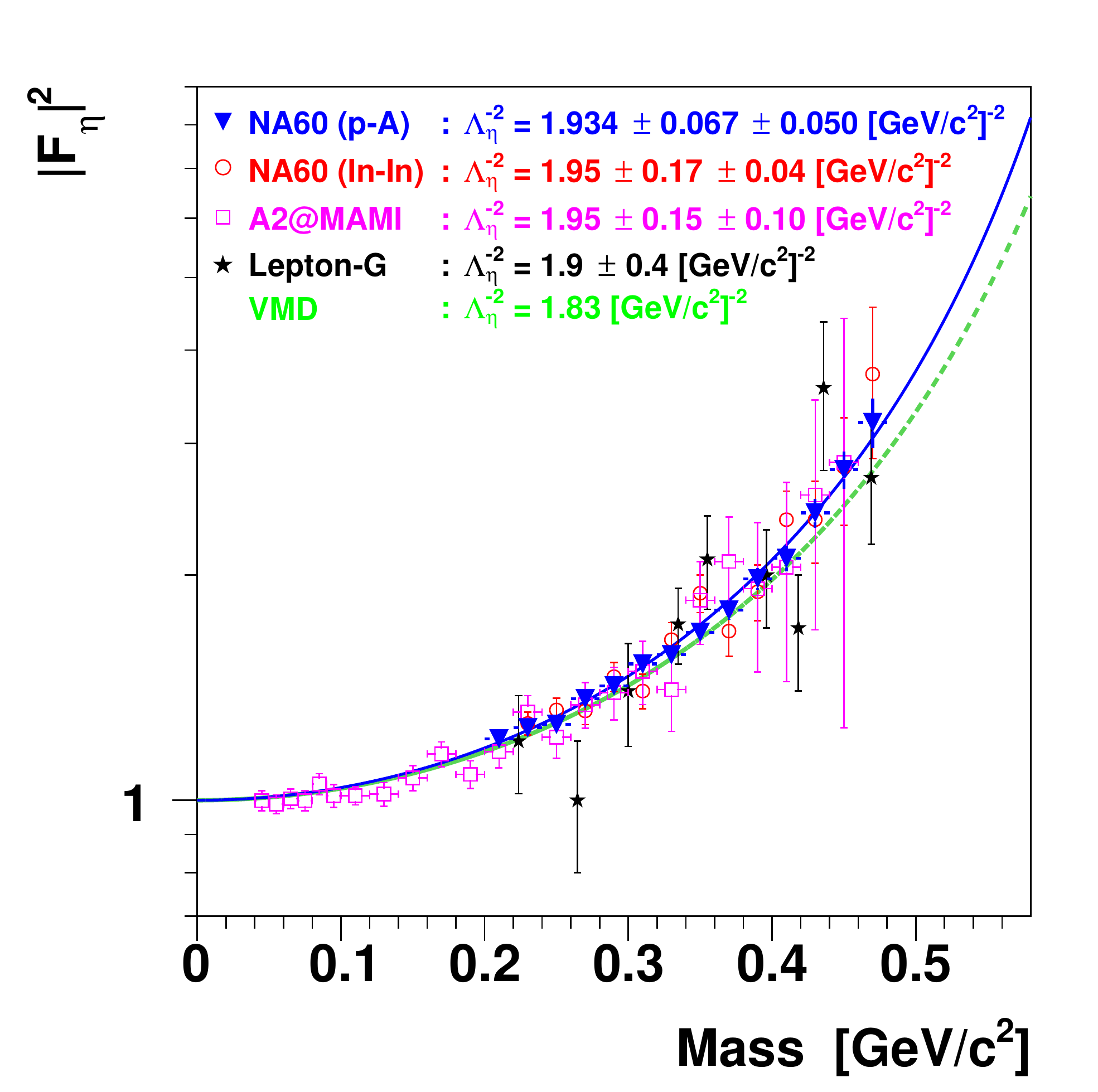}\\
    \vspace{-0.10cm}
    \hspace{-0.04\textwidth} \includegraphics[width=0.44\textwidth]{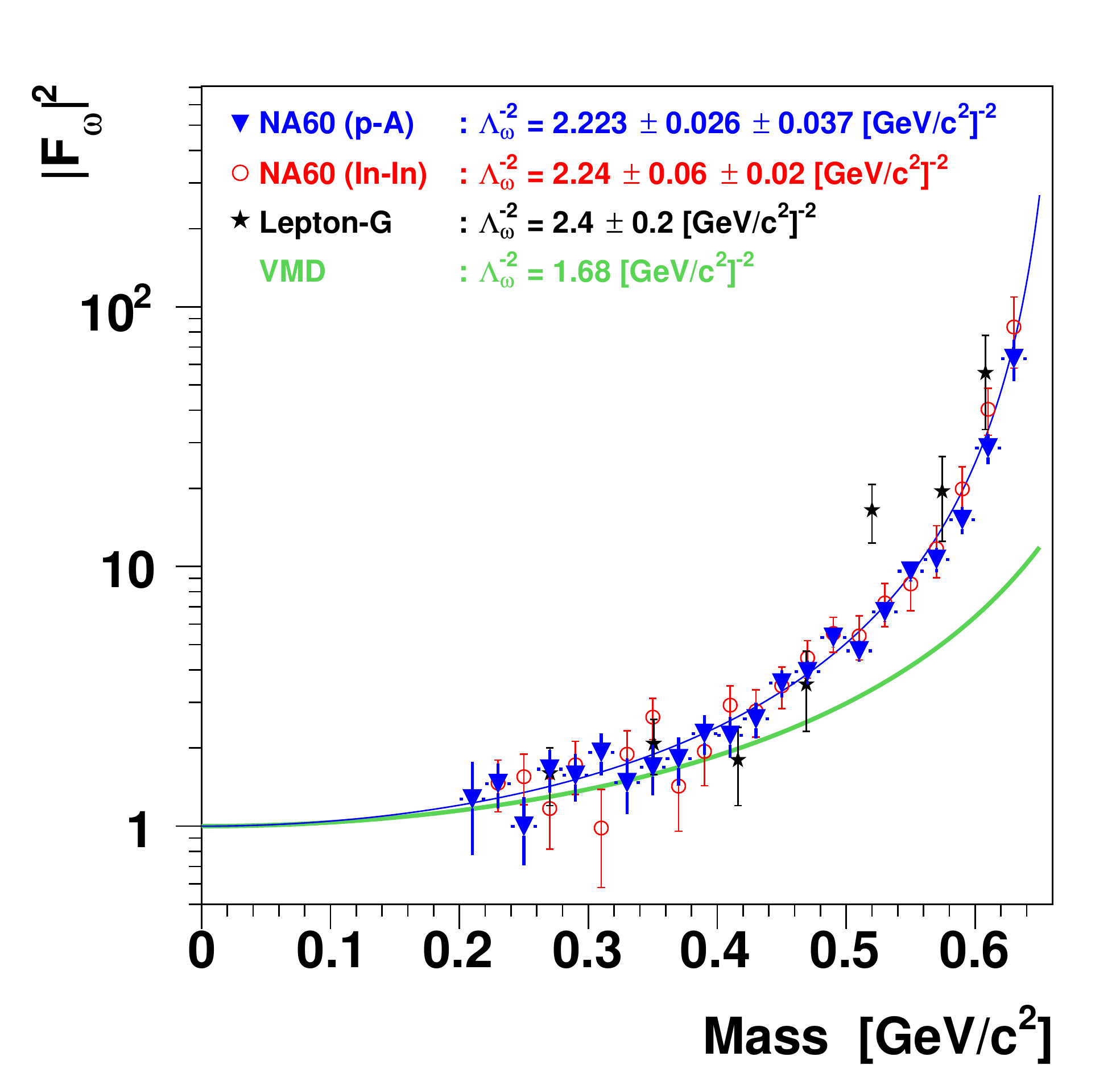}\\
    \vspace{-0.10cm}
    \hspace{-0.04\textwidth} \includegraphics[width=0.44\textwidth]{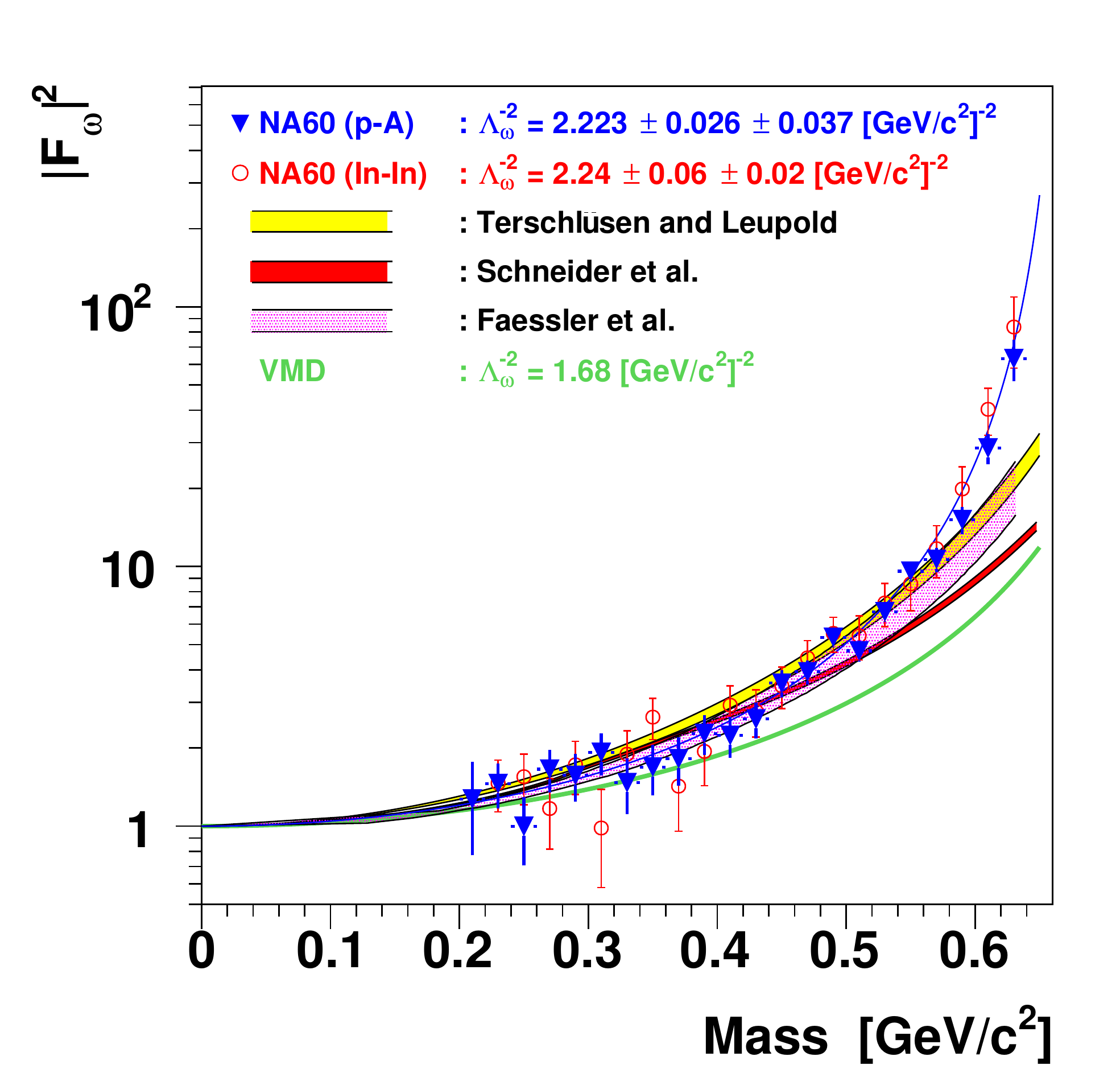}
    \end{center} 
\vspace{-0.1cm}
\caption[\textwidth]{Electromagnetic transition form factors for the
$\eta$ (top) and $\omega$ (center and bottom) mesons as a function of the dimuon mass. 
Total errors (statistical plus systematic) are associated to data points.}
\label{fig:formFactorsDisentagled}
\end{figure}

\figurename~\ref{fig:FormFactorsFit} shows the fit of the
acceptance- and efficiency-corrected mass spectrum (black triangles) with the superposition
of the processes $\eta \to \mu^+\mu^-\gamma~$,
$\omega \to \mu^+\mu^-\pi^0~$ and $\rho \to \mu^+ \mu^-$, represented by the solid lines.
In the fit, the three normalisations are left free (one for each 
of the line shapes involved) together with the
parameters $\mathrm{\Lambda}_\eta^{-2}$ and $\mathrm{\Lambda}_\omega^{-2}$, contained in
the form factors $\left| F_\eta(M^2) \right|^2$ and 
$\left|F_\omega(M^2) \right|^2$.  

Several systematic checks have been performed to test the
stability of the results and estimate their systematic
uncertainties, including the contribution from the subtraction of the known sources
from the invariant mass spectrum. They can be summarised as follows: (i) change of the
weighted acceptance by varying the $\omega$ Dalitz branching ratio
relative to $\omega\to\mu\mu$ by $\pm50\%$; (ii) change of the fit
range of the acceptance-corrected mass spectrum; (iii) change of the
$\sigma_{\eta'}/\sigma_{\omega}$ ratio including the extreme scenarios
$\sigma_{\eta'}/\sigma_{\omega} = 0$ and 4.8, covering 0\% and
400\% of the reference value $\sigma_{\eta'}/\sigma_{\omega} = 0.12$;
(iv) scaling the level of the combinatorial background by a factor between
66\% and 166\% of the reference level, fixed by the comparison with
the like-sign component of the real data; (v) considering stricter
cuts on the matching $\chi^2$ for the single muons, namely
$\chi^2_\mathrm{match}<2.5$ and $\chi^2_\mathrm{match}<2.0$ in
addition to the nominal selection $\chi^2_\mathrm{match}<3.0$.  The
normalisation of the small open charm contribution is left free, and
the fit maximises its contribution in the mass region between 1.2 and
1.4~GeV/$c^2$. In doing so, the estimation of the open charm level is
biased by the fact that the Drell-Yan process, which does not give any appreciable
contribution below 1~GeV/$c^2$ while contributing above, is neglected.
In order to study the corresponding systematic effect, we scaled the
open charm process down to 60\% and even 0\% of the level optimised by
the fit to the raw mass spectrum.  

The resulting values for the
$\mathrm{\Lambda}_\eta^{-2}$ and $\mathrm{\Lambda}_\omega^{-2}$
parameters are $1.934\ \pm\
0.067$~(stat.) $\pm\ 0.050$~(syst.)~(GeV/$c^2$)$^{-2}$ and $2.223\ \pm\ 0.026$~(stat.)
$\pm\ 0.037$~(syst.)~(GeV/$c^2$)$^{-2}$. These results are in very good agreement with
the corresponding values obtained by the analysis of the NA60
peripheral In-In data~\cite{Arnaldi:2009wb} 
$\mathrm{\Lambda}_\eta^{-2} = 1.95\ \pm\ 0.17$~(stat.) $\pm\ 0.04$~(syst.)~(GeV/$c^2$)$^{-2}$ and 
$\mathrm{\Lambda}_\omega^{-2} = 2.24\ \pm\ 0.06$~(stat.) $\pm\ 0.02$~(syst.)~(GeV/$c^2$)$^{-2}$, as well as with the Lepton-G
results~\cite{Landsberg:1986fd,Dzhelyadin:1980tj,Dzhelyadin:1980kh}. For
the form factor of the $\eta$~meson, an excellent agreement is also
found with the recent results of the A2 Collaboration at MAMI 
in the dielectron channel~\cite{Aguar-Bartolome:2013vpw}. 

Once the final fit parameters and their errors are fixed,
the contributions of the $\eta \to \mu^+\mu^-\gamma~$ and $\omega \to
\mu^+\mu^-\pi^0~$ processes are disentangled, making it possible to
present the two form factors as a function of the dimuon mass $-$ as shown in
\figurename~\ref{fig:formFactorsDisentagled}.  In a first step, we
isolate the individual Dalitz contributions in the spectrum of
\figurename~\ref{fig:FormFactorsFit}, subtracting the contribution of
the $\rho \to \mu^+ \mu^-$ decay and disentangling the $\eta \to
\mu^+\mu^-\gamma$ and $\omega \to \mu^+\mu^-\pi^0$ decays as
determined by the fit. The same individual data points are used 
for the $\eta$ and the $\omega$, subtracting for the $\eta$ the
fit results of the $\omega$ and vice versa. Since $|F_i(M)|^2 \to 1$
for $M \to 0$ by the definition of $|F_i(M)|^2$, the QED and the form
factor parts can be separately assessed for each process. This allows us,
in a second step, to obtain the squared form factors $|F_i(M)|^2$ by
dividing the data for the respective Dalitz decay by its point-like
QED part. The pole parameters and their errors as obtained from the
combined fit to both Dalitz decays (\figurename~\ref{fig:FormFactorsFit}) are
reported in \figurename~\ref{fig:formFactorsDisentagled}.
The first two panels of \figurename~\ref{fig:formFactorsDisentagled} also
include the NA60 data points obtained in
peripheral~In-In~\cite{Arnaldi:2009wb}, the Lepton-G
data~\cite{Landsberg:1986fd,Dzhelyadin:1980tj,Dzhelyadin:1980kh},
the expectations from VMD for comparison and, for the $\eta$~meson,
the data points from the recent measurement by the A2~Collaboration. Despite the much
reduced errors, the form factor of the $\eta$ is still close to the
expectation from VMD. The form factor of the $\omega$, on the other
hand, strongly deviates from the VMD expectation, showing a relative increase
close to the kinematic cut-off by a factor of $\sim10$. The corresponding data points 
are also reported in Tables~\ref{tab:FormFactor_eta} and~\ref{tab:FormFactor_omega}.

It should be noted that, in isolating the two form factors, the systematic 
uncertainties of the pole mass, the $T_\rho$ and the $\Gamma_{0\rho}$~parameters of the 
$\rho \to \mu^+ \mu^-$ line shape have been taken into account and 
properly propagated to the final points shown in 
\figurename~\ref{fig:formFactorsDisentagled}. Negligible for masses below $\sim 0.5$~GeV/$c^2$, 
this combined source of systematic uncertainty was found to be as large as one half
of the statistical uncertainty near the kinematic cut-off of the
$\omega$~form factor.
Further details on the $\rho$ line shape as used here, including the seeming absence of any 
in-medium effects on the $\rho$ distorting the shape, will be discussed in the dedicated 
Section~\ref{sec:rho} below. Other line shapes will also be discussed there, proving a remarkable
robustness of the high-mass data points of the $\omega$ form factor to
the $\rho$ line shape variations.

The extraction of the $\omega$ form factor has also been proven to be robust 
with respect to the pole parameterisation underlying the fit procedures shown 
in Figs.~\ref{fig:massSpectrum} and~\ref{fig:FormFactorsFit}. 
Since the steep mass dependence of the data points in \figurename~\ref{fig:formFactorsDisentagled} 
and its perfect description by the fit is suggestive for a pole not far above the 
kinematic limit of the Dalitz decay, an alternative parameterisation freezing the 
pole at the nominal $\rho$ position while allowing shapes very different from 
VMD was investigated as an alternative. The results were striking: satisfactory
fits to the data in \figurename~\ref{fig:FormFactorsFit} were plainly impossible, 
while the deduced form factor data points in \figurename~\ref{fig:formFactorsDisentagled} 
were absolutely immune to the bad fits in \figurename~\ref{fig:FormFactorsFit}. 



It was also specifically verified that the monotonic rise of the eta $\eta$ and $\omega$ form factors
up to the last points, close to the very steep fall-off at the kinematic limits of
their corresponding Dalitz decays, cannot, by any means,
be ascribed to an interplay between the finite mass resolution of the detector and 
the steepness of the dimuon invariant mass distribution in the considered mass regions.
The quantitative understanding of the mass resolution of the detector is best 
illustrated in the bottom panel of \figurename~\ref{fig:massSpectrum}, 
showing excellent agreement between the data and the MC 
fit in the most sensitive peak regions of the $\omega$ and the $\phi$,  
with $\chi^2$ values of about 1.2 and 1.1 averaged over the uppermost 7 points, 
respectively. 

In the bottom panel of \figurename~\ref{fig:formFactorsDisentagled},
the measured mass distribution of the muon pair in the 
$\omega \to \mu^+\mu^-\pi^0$ decay is compared to three 
recent calculations described in~\cite{Terschluesen:2010ik,Terschlusen:2012xw}, 
\cite{Schneider:2012ez} and \cite{Shekhter:2003xd,Fuchs:2005zga,fuchs}.
As can be seen, all theoretical predictions
show good agreement with the data up to $\sim 0.55$~GeV/$c^2$,
but fail to describe the data points close to the
upper kinematical boundary, $M \approx m_\omega -m_{\pi^0}$. 
Referred to the results reported in this paper, the discrepancy
is much larger than the total (statistical plus systematic) errors of the data.
Another theoretical approach
is described in~\cite{Qian:2009dc}. In this case, however, 
calculations are limited to masses below $\sim 0.4$~GeV/$c^2$ 
and the corresponding prediction is not shown here.

\begin{table}[htbp]
  \begin{center}
    \begin{tabular}{c l@{$~\pm~$}l@{$~\pm~$}l}
    \hline \hline
    \textbf{~~~~Mass~[GeV/c$\pmb{^2}$]}~~~~ & \multicolumn{3}{c}{$\pmb{|F_\eta(M)|^2}$} \\ 
    \hline \hline
    $[0.20, 0.22]$ & $ 1.208$ & $0.034$ & $0.012$\\
    $[0.22, 0.24]$ & $ 1.250$ & $0.020$ & $0.011$\\
    $[0.24, 0.26]$ & $1.264 $ & $0.022$ & $0.011$\\  
    $[0.26, 0.28]$ & $1.367 $ & $0.025$ & $0.013$\\  
    $[0.28, 0.30]$ & $1.423 $ & $0.028$ & $0.012$\\  
    $[0.30, 0.32]$ & $1.522 $ & $0.033$ & $0.014$\\  
    $[0.32, 0.34]$ & $1.565 $ & $0.038$ & $0.017$\\  
    $[0.34, 0.36]$ & $1.677 $ & $0.044$ & $0.020$\\  
    $[0.36, 0.38]$ & $1.796 $ & $0.053$ & $0.024$\\  
    $[0.38, 0.40]$ & $1.978 $ & $0.066$ & $0.028$\\  
    $[0.40, 0.42]$ & $2.105 $ & $0.083$ & $0.035$\\  
    $[0.42, 0.44]$ & $2.42 $ & $0.11  $ & $0.04  $\\  
    $[0.44, 0.46]$ & $2.77 $ & $0.16  $ & $0.07  $\\  
    $[0.46, 0.48]$ & $3.20 $ & $0.24  $ & $0.11  $\\ 
    \hline \hline
    \end{tabular}
  \caption{Electromagnetic transition form factor for the $\eta$ meson as a function of the dimuon mass. Statistical and systematic uncertainties are reported, in this order.}
  \label{tab:FormFactor_eta}
  \end{center}
\end{table}

\begin{table*}[t!]
  \begin{center}
    \begin{tabular}{c r@{$~\pm~$}l@{$~\pm~$}l c c c}
    \hline \hline
    \textbf{~~~~Mass~[GeV/c$\pmb{^2}$]}~~~~ & \multicolumn{3}{c}{$\pmb{|F_\omega(M)|^2}$} & ~~~~\textbf{Syst. Tot.}~~~~ & ~~~~\textbf{Syst. $\pmb{T_\rho}$}~~~~  &  ~~~~\textbf{Syst. $\pmb{\Gamma_\rho}$}~~~~ \\ 
    \hline \hline
   $[0.20, 0.22]$ &    1.27  &   0.50   &   0.01  &  0.9\,\%  &  0.5\,\%  & 0.5\,\%   \\
   $[0.22, 0.24]$ &    1.46  &   0.29   &   0.01  &  0.9\,\%  &  0.3\,\%  & 0.4\,\%   \\
   $[0.24, 0.26]$ &    1.00  &   0.29   &   0.01  &  0.5\,\%  &  0.4\,\%  & 0.2\,\%   \\
   $[0.26, 0.28]$ &    1.66  &   0.31   &   0.01  &  0.6\,\%  &  0.1\,\%  & 0.3\,\%   \\
   $[0.28, 0.30]$ &    1.57  &   0.33   &   0.01  &  0.4\,\%  &  0.1\,\%  & 0.2\,\%   \\
   $[0.30, 0.32]$ &    1.92  &   0.35   &   0.01  &  0.7\,\%  &  0.1\,\%  & 0.3\,\%   \\
   $[0.32, 0.34]$ &    1.47  &   0.36   &   0.01  &  0.3\,\%  &  0.2\,\%  & 0.1\,\%   \\
   $[0.34, 0.36]$ &    1.69  &   0.37   &   0.01  &  0.2\,\%  &  0.1\,\%  & 0.1\,\%   \\
   $[0.36, 0.38]$ &    1.82  &   0.38   &   0.01  &  0.4\,\%  &  0.3\,\%  & 0.2\,\%   \\
   $[0.38, 0.40]$ &    2.28  &   0.40   &   0.01  &  0.6\,\%  &  0.2\,\%  & 0.3\,\%   \\
   $[0.40, 0.42]$ &    2.24  &   0.40   &   0.02  &  0.7\,\%  &  0.5\,\%  & 0.4\,\%   \\
   $[0.42, 0.44]$ &    2.59  &   0.41   &   0.03  &  1.2\,\%  &  0.6\,\%  & 0.6\,\%   \\
   $[0.44, 0.46]$ &    3.56  &   0.42   &   0.05  &  1.5\,\%  &  0.6\,\%  & 0.8\,\%   \\
   $[0.46, 0.48]$ &    3.93  &   0.42   &   0.08  &  1.9\,\%  &  0.8\,\%  & 1.0\,\%   \\
   $[0.48, 0.50]$ &    5.33  &   0.45   &   0.12  &  2.3\,\%  &  0.8\,\%  & 1.2\,\%   \\
   $[0.50, 0.52]$ &    4.73  &   0.42   &   0.14  &  3.0\,\%  &  1.4\,\%  & 1.7\,\%   \\
   $[0.52, 0.54]$ &    6.69  &   0.54   &   0.22  &  3.2\,\%  &  1.4\,\%  & 1.8\,\%   \\
   $[0.54, 0.56]$ &    9.59  &   0.76   &   0.33  &  3.5\,\%  &  1.4\,\%  & 1.9\,\%   \\
   $[0.56, 0.58]$ &    10.6  &   1.0     &   0.47  &  4.4\,\%  &  2.0\,\%  & 2.4\,\%   \\
   $[0.58, 0.60]$ &    15.1  &   1.6     &   0.82  &  5.4\,\%  &  2.5\,\%  & 3.0\,\%   \\
   $[0.60, 0.62]$ &    28.6  &   3.4     &   1.8    &  6.2\,\%  &  2.6\,\%  & 3.4\,\%  \\
   $[0.62, 0.64]$ &    63.1  &   9.9     &   5.9    &  9.4\,\%  &  3.9\,\%  & 5.0\,\%  \\
    \hline \hline
    \end{tabular}
  \caption{Electromagnetic transition form factor for the $\omega$ meson as a function of the dimuon mass. Statistical and systematic uncertainties are reported, in this order. Contributions to the systematic
  uncertainties related to the $T_\rho$ and $\Gamma_\rho$ parameters are also separately reported.}
  \label{tab:FormFactor_omega}
  \end{center}
\end{table*}

\subsection{The $\omega \to \mu^+\mu^-\pi^0$ branching ratio}

\noindent The branching ratio for the $\omega \to \mu^+\mu^-\pi^0$
decay was measured in the same analysis, leaving free the $\omega$ Dalitz normalisation
relative to the $\omega$ two-body decay $\omega\to\mu\mu$ 
in the fit of the low mass dimuon spectrum 
(\figurename~\ref{fig:massSpectrum}).  Due to the rather small
acceptance at low $\pt$ for the $\omega$ Dalitz process, the
measurement has also been performed for $\pt > 1$~GeV/$c$ as a further
systematic check; the difference between the $\pt$-integrated value
and the value for $\pt > 1$~GeV/$c$ is $\sim 15$\%.  The final result
of the branching ratio is $BR(\omega \to \mu^+\mu^-\pi^0) = [1.41~\pm~0.09~\mathrm{(stat.)}$
$\pm~0.15~\mathrm{(syst.)}] \times 10^{-4}$. 
Within one standard deviation, this value is in agreement with that obtained in the analysis of the
peripheral In-In data: $BR(\omega \to \mu^+\mu^-\pi^0) = [1.73\ \pm\
0.25~\mathrm{(stat.)}\ \pm\ 0.14~\mathrm{(syst.)}] \times 10^{-4}$ 
and compatible with the
current value listed in the PDG~\cite{Nakamura:2010zzi} $BR(\omega \to
\mu^+\mu^-\pi^0) = (1.3~\pm~0.4) \times 10^{-4}$, which is based on
the NA60 measurement in peripheral In-In and on the older Lepton-G value.

The branching ratio $BR(\omega \to \mu^+\mu^-\pi^0)$ can alternatively be obtained by integrating
Eq.~(5) over the allowed kinematic region $2 m_\mu < M < m_\omega - m_{\pi^0}$.
This procedure, based on the knowledge of the branching ratio 
$BR(\omega \to \gamma\pi^0) = (8.28 \pm 0.28)\,\%$ according to the PDG~\cite{Nakamura:2010zzi} 
together with the present results of the omega form factor, 
leads to a value of $BR(\omega \to \mu^+\mu^-\pi^0) = (1.018 \pm 0.051) \times 10^{-4}$,
where the cited uncertainty combines the statistical and systematic uncertainties on $\mathrm{\Lambda}_\omega^{-2}$ and
the uncertainty on $BR(\omega \to \gamma\pi^0)$. There is fair agreement between the two approaches on the level
of 2~$\sigma$,  within the combined statistical and systematic uncertainties.

\subsection{$\rho$ meson line shape}
\label{sec:rho}
\noindent As already discussed, 
the value of the pole mass $m_\rho$ considered in the present
analysis was found by means of a fit to the data. 
A direct determination of the pole mass was preferred in this case,
due to the dependence of its value on the specific parameterisation chosen for the line shape.
The determination of $m_\rho$ has thus been performed
by minimising the $\chi^2$ of the fit to the low-mass spectrum, resulting in 
a pole mass of $m_\rho = (766 \pm 10)$~MeV/$c^2$. The uncertainty is the 
width of the $\chi^2$ curve around its minimum for $\mathrm{\Delta}(\chi^2/\mathrm{ndf})=1$.

The Boltzmann factor of the line shape in Eq.~(2), containing the effective temperature $T_\rho$,
is a central part of the overall phase space description. It flattens the low-mass tail, but in 
particular strongly steepens the high-mass tail, so that the slow fall-off of the broad Lorentzian 
does not continue forever.
As demonstrated by the fit to the acceptance- and efficiency-corrected mass spectrum 
in \figurename~\ref{fig:FormFactorsFit}, this factor is essential in the
description of the $\rho$ even in elementary hadronic collisions. When $T_\rho$ is left as a 
free parameter, the value $161\,\pm\,5~\mathrm{(stat.)}\,\pm\,7~\mathrm{(syst.)}$~MeV 
is found, in agreement with the
value $170\,\pm\,19~\mathrm{(stat.)}\,\pm\,3~\mathrm{(syst.)}$~MeV 
measured in peripheral In-In~\cite{Arnaldi:2009wb}. It is also consistent
with the Hagedorn temperature of 160-170~MeV, obtained by statistical model fits of
particle ratios in elementary hadron interactions and adopted in the Monte Carlo simulation
for the present analysis. This is the first measurement of
this parameter of the $\rho$ line shape in p-A collisions.

For the width $\mathrm{\Gamma}_{0\rho}$, an optimised value of
$\mathrm{\Gamma}_{0\rho} = 146 \pm 6~\mathrm{(stat.)}$~MeV was found when leaving this
parameter free in the fit. This value is compatible with the PDG
one~\cite{Nakamura:2010zzi} $\mathrm{\Gamma}_{0\rho} = 149$~MeV
considered in the rest of the analysis.

The perfect fit of the line shape Eq.~(2) as visible in \figurename~\ref{fig:FormFactorsFit} raises the question on the seeming
absence of any noticeable broadening of the $\rho$ by in-medium effects within the given errors.
While such effects strongly appear in A-A collisions, due to the creation of a hot and dense medium
embedding the rho, the situation for p-A interactions is very sensitive to the beam-energy scale. 
While at energies of the order of a few GeV cold nuclear matter effects do
exist, p-A interactions at 400~GeV
are expected to be essentially free from them, on simple kinematic grounds. The leading
proton in the laboratory frame has a rapidity of about~6, while the
cold target nucleus is left behind at 
rapidities around~0, except for a few hit nucleons tailing up to at most mid-rapidity. Therefore,
there is no cold medium to speak of at mid-rapidity, where particle production is measured in
this experiment (about 3-4 in the laboratory frame). On the other hand, the rapidity density of the produced particles
is not much higher than in genuine pp in this region. The absence of sizeable in-medium effects
under the conditions of the present experiment is therefore hardly surprising.

To shed further independent light on the sensitivity to in-medium effects of the $\rho$ in the present
experiment, the $\rho$ line shapes measured at much lower energies --- by the
CLAS experiment at JLab in $\gamma$-A up to 4~GeV~\cite{Nasseripour:2007mga}
and by the KEK-PS E325 experiment in p-A at 12~GeV~\cite{Naruki:2005kd}, and published as (mutually contradicting) 
evidence for in-medium effects --- were also used in the fits to the present data. The description 
was equally unacceptable for both options, with data-MC residuals far
outside the data errors and the $\chi^2/\mathrm{ndf}$ found to be as large
as $\sim 6$
and $\sim 3$ for the CLAS and KEK line shapes. This sets
an independent quantitative limit on possible in-medium effects of the
$\rho$ in ultra-relativistic
p-A collisions: it is far below the level observed at JLab and KEK energies. At the same time, 
the influence of assuming the low-energy shapes for the $\rho$ meson
when extracting the form factor data of the $\omega$ Dalitz 
decay in the relevant mass region above 0.55~GeV/$c^2$ was found to be at most at the edge of the errors 
shown in Figure~3, emphasising a remarkable robustness of those data points even in this most 
sensitive region.    
 

\subsection{$\rho/\omega$ interference}
\label{sec:interference}

\noindent In the analysis described up to now, the 
$\rho$ and $\omega$ contributions are added incoherently to the MC 
cocktail describing the low mass dimuon spectrum.
However, in the presence of common production mechanisms for the $\rho$ and $\omega$ mesons, 
quantum interference effects may occur in a decay channel common to the two particles $-$ for instance 
$e^+e^-$ or $\mu^+\mu^-$. In this case, the interference line shape can be described 
as the coherent sum of the $\rho$ and $\omega$ amplitudes
\begin{equation}
\big|A_{\rho+\omega}\big|^2\propto \big|\tilde F_\rho(M) + R \tilde F_\omega(M) \big|^2
 \times \big| A_{\gamma^{*}\to \mu^+\mu^-}\big|^2,
 \label{eq:interference}
\end{equation}
where $\tilde F_{\rho,\omega}(M)$ are the normalised $\rho$ and $\omega$ propagators, 
$F_{\rho,\omega}(M)=1/(M^2-m^2_{\rho,\omega}+im_{\rho,\omega}\mathrm{\Gamma}_{\rho,\omega})$ and $R$ is a complex parameter.
In order to study the sensitivity of the data to any possible interference effect
between the $\rho$ and the $\omega$ mesons, the analysis was repeated summing the
$\rho$ and $\omega$ amplitudes coherently as described by Eq.~(\ref{eq:interference}), with
the complex parameter $R$ expressed as $R=|R|e^{i\alpha}$.

\begin{figure}[ht] 
   \begin{center}
    \includegraphics[width=0.40\textwidth]{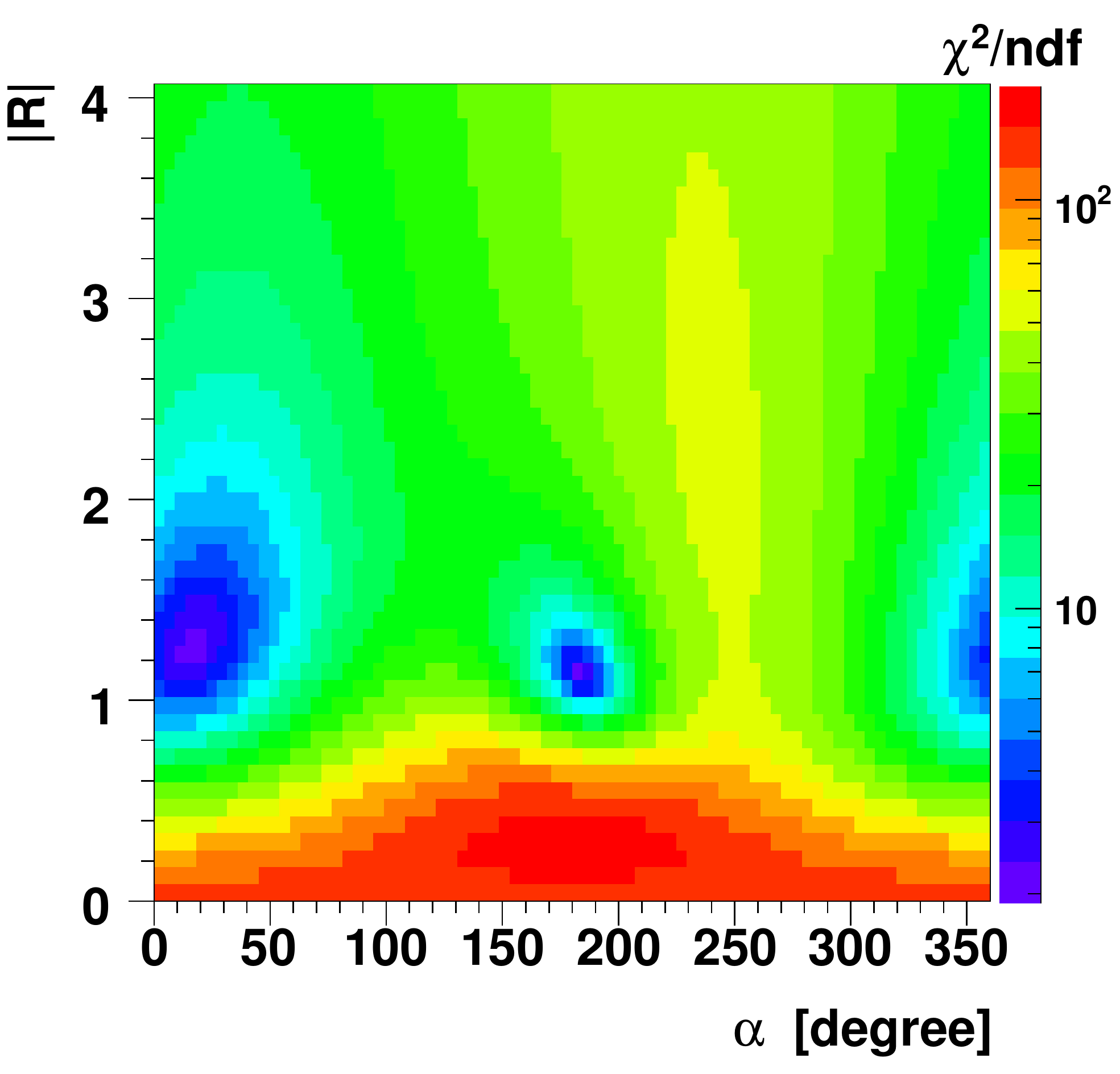}
    \end{center} 
    \vspace{-0.3cm}
\caption[\textwidth]{Fit $\chi^2/\mathrm{ndf}$ for coherent fits as a function of $|R|$ and $\alpha$.}
\label{fig:interference}
\vspace{0.5cm}
\end{figure}

In an analysis in which the $\rho$ and $\omega$ are summed coherently, one has to find the
best values for $|R|$ and $\alpha$, by minimising the $\chi^2$ of the fit to the low mass spectrum.
The resulting $\chi^2/\mathrm{ndf}$ as a function of $|R|$ and $\alpha$ is shown in \figurename~\ref{fig:interference}.
As seen from this $\chi^2$ map, two minima are present. The statistical errors of 
the parameters corresponding to the minima
are obtained considering the region defined by $\mathrm{\Delta}(\chi^2/\mathrm{ndf})=1$.
In this way one obtains $|R|=1.27 \pm 0.17$, $\alpha = 15^\circ \pm 15^\circ$ for the first minimum, 
and $|R| = 1.10 \pm 0.08$, $\alpha = 180^\circ \pm 5^\circ$ for the second. \\

\noindent The first minimum corresponds to constructive interference, while the second
to destructive interference. The data seem to rule out $\rho/\omega$ interference scenarios other than 
completely constructive or destructive. It should be noted that, 
given the experimental mass resolution and data errors,
these interference scenarios give equivalent descriptions of the data. 
No significant difference was found. Given that a satisfactory description of
the data can be obtained also with the incoherent sum of $\rho$ and $\omega$ contributions,
we cannot make any statement on the existence of interference with parameters inside the
two mentioned regions, but we can exclude it outside of these regions.

Two previous experiments tried to assess the $\rho/\omega$ interference effect in p-A
collisions. The HELIOS/I experiment studied p-Be collisions at~450 GeV and
found $\alpha=100^\circ \pm 30^\circ$~\cite{Veenhof:1993xt,Akesson:1994mb}.
The already cited KEK-PS E325 experiment~\cite{Naruki:2005kd} studied the $e^+e^-$ decay channel in p-C and p-Cu 
collisions at 12~GeV and, in the attempt to describe the $\rho/\omega$ peak 
with an interference pattern, reported $\alpha \simeq 160^\circ$ (no error was quoted). 
However, in that case the interference effect was finally disfavored 
to possible evidence for a mass shift of the $\rho$ meson.

\section{Conclusions}

\noindent A detailed analysis of the low mass dimuon data collected by NA60 in p-A 
collisions at 400~GeV has been performed. The large sample of high quality
data allowed a new measurement of the electromagnetic transition form
factors of the $\eta$ and $\omega$ mesons, improving by a factor~3 the precision of
the previous measurement, made by NA60 in peripheral In-In collisions. The new results
presented here confirm on more solid ground the discrepancy between the available predictions for the  
form factor of the $\omega$ meson and the experimental data
close to the kinematic limit. The same analysis also allowed 
an improved measurement of the branching ratio of the Dalitz decay
$\omega \to \mu^+\mu^-\pi^0$.

The $\rho$ line shape has also been studied in detail, confirming the importance of the
Boltzmann factor, for which a measure of the $T_\rho$ parameter has been performed,
for the first time in p-A collisions.
The existence of a possible $\rho/\omega$ interference effect has been investigated, 
ruling out interference scenarios other than completely constructive or destructive.
The residual ambiguity between constructive interference, destructive interference and 
incoherent superposition of $\rho$ and $\omega$, with equivalent descriptions of the data,
could not be resolved. Given the overall quality of the comparison between data and 
expected sources, no evidence is found for in-medium cold nuclear matter effects in the
$\rho/\omega$ region. \\

~\\
\noindent The authors are very grateful to B.~Friman for his concise clarification of
polarisation in Dalitz decays and helpful discussions on $\rho/\omega$ interference.

\bibliographystyle{utphys}
\bibliography{na60_pA2004_form-factors}

\end{document}